\documentclass[12pt]{article}
\usepackage{amssymb,amsmath}
\usepackage{hyperref}
\usepackage{graphicx}
\usepackage{color}

\begin{document}

\title{\LARGE\bf A rational approximation for efficient computation of the Voigt function in quantitative spectroscopy}

\author{
\normalsize\bf S. M. Abrarov\footnote{\scriptsize{Dept. Earth and Space Science and Engineering, York University, Toronto, Canada, M3J 1P3.}}\, and B. M. Quine$^{*}$\footnote{\scriptsize{Dept. Physics and Astronomy, York University, Toronto, Canada, M3J 1P3.}}}

\date{March, 27 2015}
\maketitle

\begin{abstract}
We present a rational approximation for rapid and accurate computation of the Voigt function, obtained by residue calculus. The computational test reveals that with only $16$ summation terms this approximation provides average accuracy ${10^{ - 14}}$ over a wide domain of practical interest  $0 < x < 40,000$ and ${10^{ - 4}} < y < {10^2}$ for applications using the HITRAN molecular spectroscopic database. The proposed rational approximation takes less than half the computation time of that required by Weideman\text{'}s rational approximation. Algorithmic stability is achieved due to absence of the poles at $y \geqslant 0$ and $ - \infty  < x < \infty $.
\vspace{0.25cm}
\vspace{0.25cm}
\end{abstract}

\section{Introduction}

The Voigt function is widely used and finds broad applications in many scientific disciplines \cite{Armstrong1967, Armstrong1972, Schreier1992, Letchworth2007, Pagnini2010}. It is commonly applied in Applied Mathematics, Physics, Chemistry and Astronomy as it describes the line profile behavior that occurs due to simultaneous Lorentz and Doppler broadening effects; the Lorentz broadening is observed as a result of the Heisenberg uncertainty principle and chaotic multiple collisions of the particles while the Doppler broadening appears due to velocity distribution of the particles.

The Voigt function can describe the spectral properties in the photon emission or absorption of atmospheric gases \cite{Edwards1992, Quine2002, Christensen2012, Berk2013, Quine2013} and celestial bodies \cite{Emerson1996}. It is also widely used in crystallography \cite{Prince2004} and can be utilized in many other spectroscopic applications, for example, to characterize the photo-luminescent properties of nanomaterials \cite{Miyauchi2013} or to determine the hyper structure of an isotope \cite{Sonnenschein2012} and so on.

Mathematically, the Voigt function is a convolution integral of the Cauchy and Gaussian distributions \cite{Armstrong1967, Armstrong1972, Schreier1992, Letchworth2007, Pagnini2010}
\begin{equation}\label{eq_1}
K\left( {x,y} \right) = \frac{y}{\pi }\int\limits_{ - \infty }^\infty  {\frac{{{e^{ - {t^2}}}}}{{{y^2} + {{\left( {x - t} \right)}^2}}}dt}
\end{equation}
and represents the real part of the complex probability function \cite{Armstrong1972, Schreier1992}
\begin{equation}\label{eq_2}
W\left( z \right) = \frac{i}{\pi }\int\limits_{ - \infty }^\infty  {\frac{{{e^{ - {t^2}}}}}{{z - t}}dt}
\end{equation}
where $z = x + iy$ is a complex argument. The complex probability function can be expressed explicitly as a superposition of the real and imaginary parts $W\left( {x,y} \right) = K\left( {x,y} \right) + iL\left( {x,y} \right)$,
where its imaginary part is given by \cite{Armstrong1972, Schreier1992}
\begin{equation}\label{eq_3}
L\left( {x,y} \right) = \frac{1}{\pi }\int\limits_{ - \infty }^\infty  {\frac{{\left( {x - t} \right){e^{ - {t^2}}}}}{{{y^2} + {{\left( {x - t} \right)}^2}}}dt}.
\end{equation}

Another closely related function is the complex error function, also known as the Faddeeva function \cite{Schreier1992, Faddeyeva1961, Gautschi1970, Abramowitz1972, Poppe1990a, Weideman1994}
\begin{equation}\label{eq_4}
\begin{aligned}
  w\left( z \right) &= {e^{ - {z^2}}}\left[ {1 - {\text{erf}}\left( { - iz} \right)} \right] \\ 
   &= {e^{ - {z^2}}}\left( {1 + \frac{{2i}}{{\sqrt \pi  }}\int\limits_0^z {{e^{{t^2}}}} dt} \right),
\end{aligned}
\end{equation}

There is a relation between complex probability function \eqref{eq_2} and complex error function \eqref{eq_4}
$$
W\left( z \right) = w\left( z \right), \quad\quad\quad \operatorname{Im} \left[ z \right] \geqslant 0
$$
or
\begin{equation}\label{eq_5}
w\left( {x,y} \right) = \underbrace {K\left( {x,y} \right) + iL\left( {x,y} \right)}_{W\left( {x,y} \right)}, \quad\quad y \geqslant 0.
\end{equation}

In order to describe spectral characteristics of a system with high resolution, intense computation is required. For example, in a line-by-line radiative transfer simulation to resolve some problems associated with inhomogeneity, the Earth\text{'}s or other planetary atmosphere can be divided up to $1000$ layers \cite{Edwards1992, Quine2002}. Taking into account that computation requires a nested loop procedure in order to adjust properly for the fitting parameters for each atmospheric layer that may contain many different molecular species, the total number of the computed points may exceed hundreds of millions. Since in a radiative transfer model the computation of spectral broadening profiles requires considerable amount of time, a rapid approximation of the Voigt function is very desirable \cite{Edwards1992, Quine2002}. Consequently, the rapid and accurate computation of the Voigt/complex error function still remains topical (see for example an optimized algorithm in the recent work \cite{Karbach2014}).

In this work we present a new rational approximation of the Voigt function for efficient computation. Due to absence of the poles at $y \geqslant 0$ and $ - \infty  < x < \infty $ this rational approximation enables stability in algorithmic implementation.

\section{Derivation of the rational approximation}

The complex error function \eqref{eq_4} can also be expressed in an alternative form as \cite{Srivastava1987, Abrarov2011}
$$
w\left( {x,y} \right) = \frac{1}{{\sqrt \pi  }}\int\limits_0^\infty  {\exp \left( { - {t^2}/4} \right)\exp \left( { - yt} \right)\exp \left( {ixt} \right)dt},
$$
where its real and imaginary parts are
$$
\operatorname{Re} \left[ {w\left( {x,y} \right)} \right] = \frac{1}{{\sqrt \pi  }}\int\limits_0^\infty  {\exp \left( { - {t^2}/4} \right)\exp \left( { - yt} \right)\cos \left( {xt} \right)dt}
$$
and 
$$
\operatorname{Im} \left[ {w\left( {x,y} \right)} \right] = \frac{1}{{\sqrt \pi  }}\int\limits_0^\infty  {\exp \left( { - {t^2}/4} \right)\exp \left( { - yt} \right)\sin \left( {xt} \right)dt},
$$
respectively. By changing sign of the variable $x$ to negative in the last two equations above, we can see the symmetric properties of the complex error function
$$
\left\{ \begin{aligned}
&{\mathop{\rm Re}\nolimits} \left[ {w\left( {x,y} \right)} \right] = {\mathop{\rm Re}\nolimits} \left[ {w\left( { - x,y} \right)} \right]\\
&{\mathop{\rm Im}\nolimits} \left[ {w\left( { - x,y} \right)} \right] =  - {\mathop{\rm Im}\nolimits} \left[ {w\left( {x,y} \right)} \right].
\end{aligned} \right.
$$
Consequently, it follows that
\begin{equation}\label{eq_6}
\operatorname{Re} \left[ {w\left( {x,y} \right)} \right] = \left[ {w\left( {x,y} \right) + w\left( { - x,y} \right)} \right]/2
\end{equation}
and
$$
\operatorname{Im} \left[ {w\left( {x,y} \right)} \right] = \left[ {w\left( {x,y} \right) - w\left( { - x,y} \right)} \right]/2.
$$
It is worth noting that with these identities and equation \eqref{eq_4} we can also obtain two interesting relations for the real and imaginary parts of the error function of complex argument as follows
$$
\left\{ \begin{aligned}
{\mathop{\rm Re}\nolimits} \left[ {{\rm{erf}}\left( {x + iy} \right)} \right] &= \frac{{{\rm{erf}}\left( {x + iy} \right) + {\rm{erf}}\left( {x - iy} \right)}}{2}\\
{\mathop{\rm Im}\nolimits} \left[ {{\rm{erf}}\left( {x + iy} \right)} \right] &= \frac{{{\rm{erf}}\left( {x + iy} \right) - {\rm{erf}}\left( {x - iy} \right)}}{{2i}}.
\end{aligned} \right.
$$

Since \cite{Armstrong1967}
$$
\mathop {\lim }\limits_{y \to 0} \frac{y}{{{y^2} + {{\left( {x - t} \right)}^2}}}{e^{ - {t^2}}} = \pi \delta \left( {x - t} \right){e^{ - {t^2}}},
$$
where $\delta \left( {x - t} \right)$ is the Dirac\text{'}s delta function, we obtain
$$
\begin{aligned}
{\left. {\frac{y}{\pi }\int\limits_{ - \infty }^\infty  {\frac{{{e^{ - {t^2}}}}}{{{y^2} + {{\left( {x - t} \right)}^2}}}dt} } \right|_{y = 0}}
& = \mathop {\lim }\limits_{y \to 0} \frac{1}{\pi }\int\limits_{ - \infty }^\infty  {\frac{{y{e^{ - {t^2}}}}}{{{y^2} + {{\left( {x - t} \right)}^2}}}dt} \\
& = \frac{1}{\pi }\int\limits_{ - \infty }^\infty  {\pi \delta \left( {x - t} \right){e^{ - {t^2}}}dt}  = {e^{ - {x^2}}}.
\end{aligned}
$$
Consequently, we can write
${\mathop{\rm Re}\nolimits} \left[ {w\left( {x,y = 0} \right)} \right] = K\left( {x,y = 0} \right) \equiv \exp \left( { - {x^2}} \right)$
and from the identity \eqref{eq_6} it immediately follows that
\begin{equation}\label{eq_7}
\exp \left( { - {x^2}} \right) = \left[ {K\left( {x,y = 0} \right) + K\left( { - x,y = 0} \right)} \right]/2.
\end{equation}

In our recent publication we have shown that a sampling methodology based on incomplete expansion of the sinc function leads to a new series approximation of the complex error function \cite{Abrarov2015b}
\begin{equation}\label{eq_8}
w\left( z \right) = W\left( z \right) \approx \sum\limits_{m = 1}^{{2^{M - 1}}} {\frac{{{A_m} + \left( {z + i\varsigma /2} \right){B_m}}}{{C_m^2 - {{\left( {z + i\varsigma /2} \right)}^2}}}}, \quad\quad \operatorname{Im} \left[ z \right] \geqslant 0.
\end{equation}
where the coefficients are
$$
{A_m} = \frac{{\sqrt \pi  \left( {2m - 1} \right)}}{{{2^{2M}}h}}\sum\limits_{n =  - N}^N {{e^{{\varsigma ^2}/4 - {n^2}{h^2}}}\sin \left( {\frac{{\pi \left( {2m - 1} \right)\left( {nh + \varsigma /2} \right)}}{{{2^M}h}}} \right)},
$$
$$
{B_m} =  - \frac{i}{{{2^{M - 1}}\sqrt \pi  }}\sum\limits_{n =  - N}^N {{e^{{\varsigma ^2}/4 - {n^2}{h^2}}}\cos \left( {\frac{{\pi \left( {2m - 1} \right)\left( {nh + \varsigma /2} \right)}}{{{2^M}h}}} \right)},
$$
$$
{C_m} = \frac{{\pi \left( {2m - 1} \right)}}{{{2^{M + 1}}h}}
$$
with $\varsigma  = 2.75$, $h = 0.25$, $M = 5$ and $N = 23$. As we can see, the integer on upper limit of the summation in this approximation is equal to ${2^{M - 1}}$. However, this restriction can be omitted and application of the series approximation above can be generalized for any integer.

Consider the following limit for the sinc function \cite{Abrarov2015b}
\small
$$
{\text{sinc}}\left( t \right) = \mathop {\lim }\limits_{M \to \infty } \frac{1}{{{2^{M - 1}}}}\sum\limits_{m = 1}^{{2^{M - 1}}} {\cos \left( {\frac{{2m - 1}}{{{2^M}}}t} \right)}  = \mathop {\lim }\limits_{M \to \infty } \frac{1}{{{2^{M - 1}}}}\sum\limits_{m = 1}^{{2^{M - 1}}} {\cos \left( {\frac{{m - 1/2}}{{{2^{M - 1}}}}t} \right)},
$$
\normalsize
where we imply that the sinc function is defined as 
$$
\left\{ {{\text{sinc}}\left( {t \ne 0} \right) = \sin \left( t \right)/t,\,\,{\text{sinc}}\left( {t = 0} \right) = 1} \right\}.
$$
Change of the integer variable ${2^{M - 1}} \to {m_{\max }}$ in this limit leads to
$$
{\text{sinc}}\left( t \right) = \mathop {\lim }\limits_{{m_{\max }} \to \infty } \frac{1}{{{m_{\max }}}}\sum\limits_{m = 1}^{{m_{\max }}} {\cos \left( {\frac{{m - 1/2}}{{{m_{\max }}}}t} \right)}.
$$
This signifies that if the integer ${m_{\max }}$ is large enough, it retains all properties required to approximate the sinc function that can be used for sampling (see sampling methodology in \cite{Abrarov2015b} for details). Consequently, we can generalize the approximation \eqref{eq_8} of the complex error function for an arbitrary integer ${m_{\max }}$ as follows
\begin{equation}\label{eq_9}
w\left( z \right) = W\left( z \right) \approx \sum\limits_{m = 1}^{{m_{\max }}} {\frac{{{A_m} + \left( {z + i\varsigma /2} \right){B_m}}}{{C_m^2 - {{\left( {z + i\varsigma /2} \right)}^2}}}}, \quad\quad \operatorname{Im} \left[ z \right] \geqslant 0.
\end{equation}
where the corresponding coefficients are rewritten as
$$
{A_m} = \frac{{\sqrt \pi  \left( {m - 1/2} \right)}}{{2m_{\max }^2h}}\sum\limits_{n =  - N}^N {{e^{{\varsigma ^2}/4 - {n^2}{h^2}}}\sin \left( {\frac{{\pi \left( {m - 1/2} \right)\left( {nh + \varsigma /2} \right)}}{{{m_{\max }}h}}} \right)},
$$
$$
{B_m} =  - \frac{i}{{{m_{\max }}\sqrt \pi  }}\sum\limits_{n =  - N}^N {{e^{{\varsigma ^2}/4 - {n^2}{h^2}}}\cos \left( {\frac{{\pi \left( {m - 1/2} \right)\left( {nh + \varsigma /2} \right)}}{{{m_{\max }}h}}} \right)},
$$
$$
{C_m} = \frac{{\pi \left( {m - 1/2} \right)}}{{2{m_{\max }}h}}.
$$

Combining identity \eqref{eq_7} and approximation \eqref{eq_9} together at $y = 0$ yields an exponential function approximation
$$
\exp \left( { - {x^2}} \right) \approx \frac{1}{2}\sum\limits_{m = 1}^{{m_{\max }}} {\left[ {\frac{{{A_m} + \left( {x + i\varsigma /2} \right){B_m}}}{{C_m^2 - {{\left( {x + i\varsigma /2} \right)}^2}}} + \frac{{{A_m} + \left( { - x + i\varsigma /2} \right){B_m}}}{{C_m^2 - {{\left( { - x + i\varsigma /2} \right)}^2}}}} \right]}.
$$

Figure 1 shows the difference $\varepsilon \left( t \right)$ between the original exponential function $\exp \left( { - {t^2}} \right)$ and its approximation
\[
\varepsilon \left( t \right) = \exp \left( { - {t^2}} \right) - \frac{1}{2}\sum\limits_{m = 1}^{{m_{\max }}} {\left[ {\frac{{{A_m} + \left( {x + i\varsigma /2} \right){B_m}}}{{C_m^2 - {{\left( {x + i\varsigma /2} \right)}^2}}} + \frac{{{A_m} + \left( { - x + i\varsigma /2} \right){B_m}}}{{C_m^2 - {{\left( { - x + i\varsigma /2} \right)}^2}}}} \right]}.
\]
As we can see from this figure, even with only ${m_{\max }} = 16$ summation terms the difference $\varepsilon \left( t \right)$ is very small and remains within the narrow range $ \pm 5 \times {10^{ - 10}}$. This confirms a rapid convergence of the exponential function approximation that makes it suitable for numerical integration. Specifically, this series approximation can be further used to replace the original exponential function $\exp \left( { - {t^2}} \right)$ from the integrand in integral equation \eqref{eq_1} as follows
\footnotesize
\begin{equation}\label{eq_10}
\begin{aligned}
&K\left( {x,y} \right) \approx \\ 
&\frac{y}{{2\pi }}\int\limits_{ - \infty }^\infty  {\frac{1}{{{y^2} + {{\left( {x - t} \right)}^2}}}\sum\limits_{m = 1}^{{m_{\max }}} {\left[ {\frac{{{A_m} + \left( {t + i\varsigma /2} \right){B_m}}}{{C_m^2 - {{\left( {t + i\varsigma /2} \right)}^2}}} + \frac{{{A_m} + \left( { - t + i\varsigma /2} \right){B_m}}}{{C_m^2 - {{\left( { - t + i\varsigma /2} \right)}^2}}}} \right]} } \,dt,
\end{aligned}
\end{equation}
\normalsize

\begin{figure}[ht]
\begin{center}
\includegraphics[width=24pc]{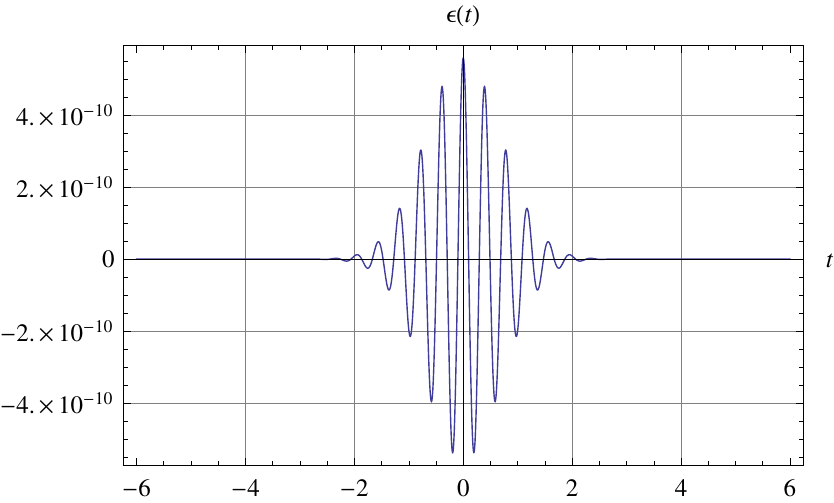}\hspace{2pc}%
\begin{minipage}[b]{28pc}
\vspace{0.3cm}
{\sffamily {\bf{Fig. 1.}} The difference $\varepsilon \left( t \right)$ between the original exponential function $\exp \left( { - {t^2}} \right)$ and its approximation at ${m_{\max }} = 16$.}
\end{minipage}
\end{center}
\end{figure}

Consider the series approximation \eqref{eq_10} of the Voigt function in more detail. The integrand in this integral is analytic everywhere over the entire complex plain except $2 + 4 m_{\text{max}}$ isolated points
\small
$$
\begin{aligned}
\left\{ x - iy,x + iy, - {C_m} - i\varsigma /2,{C_m} - i\varsigma /2, - {C_m} + i\varsigma /2,{C_m} \right. &+ \left. i\varsigma /2 \right\}, \\
m &\in \left\{ {1,2,3,\,\, \ldots \,\,m_{\text{max}}} \right\}
\end{aligned}
$$
\normalsize
where singularities are observed. However, as we take a contour integral only on the upper complex plane, for example as a semicircle ${C_{ccw}}$ with infinite radius in counterclockwise (CCW) direction, the quantity of isolated points is reduced twice and becomes equal to $1 + 2 m_{\text{max}}$.

Lastly, substituting the corresponding isolated points inside the domain enclosed by contour ${C_{ccw}}$:
\[
{t_r} = \left\{ {x + iy, - {C_m} + i\varsigma /2,{C_m} + i\varsigma /2} \right\}, \quad\quad m \in \left\{ {1,2,3,\,\, \ldots \,\,m_{\text{max}}} \right\}
\]
into the Residue Theorem\text{'}s formula that for our specific case is expressed in form
\[
\frac{1}{{2\pi i}}\oint\limits_{{C_{ccw}}} {f\left( t \right)} \,dt = \sum\limits_{r = 1}^{1 + 2 m_{\text{max}}} {{\text{Res}}\left[ {f\left( t \right),{t_r}} \right]},
\]
where $f\left( t \right)$ is the integrand of integral \eqref{eq_10}, we find a new series approximation of the Voigt function
\footnotesize
\begin{equation}\label{eq_11}
\begin{aligned}
&K\left( {x,y} \right) \approx \\
&\sum\limits_{m = 1}^{{m_{\max }}} {\frac{{{A_m}\left[ {C_m^2 - {x^2} + {{\left( {y + \varsigma /2} \right)}^2}} \right] + i{B_m}\left( {y + \varsigma /2} \right)\left[ {C_m^2 + {x^2} + {{\left( {y + \varsigma /2} \right)}^2}} \right]}}{{\left[ {{C_m} + x - i\left( {y + \varsigma /2} \right)} \right]\left[ {{C_m} - x + i\left( {y + \varsigma /2} \right)} \right]\left[ {C_m^2 - {{\left( {x + i\left( {y + \varsigma /2} \right)} \right)}^2}} \right]}}}.
\end{aligned}
\end{equation}
\normalsize
Since an algorithm involving complex numbers takes extra time, it would be very desirable to exclude them in computation. Thus, after some trivial rearrangements of the equation above, it can be represented in a simplified form as the series approximation consisting of the real variables and constants only
\begin{equation}\label{eq_12}
\begin{aligned}
\kappa \left( {x,y} \right) &\triangleq \sum\limits_{m = 1}^{{m_{\max }}} {\frac{{{\alpha _m}\left( {{\beta _m} + {y^2} - {x^2}} \right) + {\gamma _m}y\left( {{\beta _m} + {x^2} + {y^2}} \right)}}{{\beta _m^2 + 2{\beta _m}\left( {{y^2} - {x^2}} \right) + {{\left( {{x^2} + {y^2}} \right)}^2}}}} \\
   &\Rightarrow K\left( {x,y} \right) \approx \kappa \left( {x,y + \varsigma /2} \right),
\end{aligned}
\end{equation}
where
$$
{\alpha _m} = {A_m} = \frac{{\sqrt \pi  \left( {m - 1/2} \right)}}{{2m_{\max }^2h}}\sum\limits_{n =  - N}^N {{e^{{\varsigma ^2}/4 - {n^2}{h^2}}}\sin \left( {\frac{{\pi \left( {m - 1/2} \right)\left( {nh + \varsigma /2} \right)}}{{{m_{\max }}h}}} \right)},
$$
$$
{\beta _m} = C_m^2 = {\left( {\frac{{\pi \left( {m - 1/2} \right)}}{{2{m_{\max }}h}}} \right)^2}
$$
and
$$
{\gamma _m} = i{B_m} = \frac{1}{{{m_{\max }}\sqrt \pi  }}\sum\limits_{n =  - N}^N {{e^{{\varsigma ^2}/4 - {n^2}{h^2}}}\cos \left( {\frac{{\pi \left( {m - 1/2} \right)\left( {nh + \varsigma /2} \right)}}{{{m_{\max }}h}}} \right)}.
$$
As the constants ${\alpha _m}$, ${\beta _m}$ and ${\gamma _m}$ are independent of the input parameters $x$ and $y$, the obtained series \eqref{eq_12} is a rational approximation.

\section{Implementation}

Since the Voigt function is an even with respect to the parameter $x$ and odd with respect to the parameter $y$:
$$
K\left( {x, - \left| y \right|} \right) = K\left( { - x, - \left| y \right|} \right) =  - K\left( {x,\left| y \right|} \right),
$$
it is sufficient to consider the values $x$ and $y$ only from the ${{\text{I}}^{{\text{st}}}}$ and ${\text{I}}{{\text{I}}^{{\text{nd}}}}$ quadrants in order to cover the entire complex plane. Consequently, in algorithmic implementation it is reasonable to take the second input parameter by modulus as $\left| y \right|$ and compute the Voigt function according to the scheme
$$
\left\{ \begin{aligned} 
&K\left( {x,y > 0} \right) \approx \kappa \left( {x,\left| y \right| + \varsigma /2} \right) \\ 
&K\left( {x,y < 0} \right) \approx  - \kappa \left( {x,\left| y \right| + \varsigma /2} \right) 
\end{aligned}  \right.
$$
Thus, if the parameter $y$ is negative, we first take it by absolute value and, after computation, simply change the sign of the computed result to opposite. It should also be noted that taking the argument $\left| y \right|$ is advantageous in implementation as it prevents computational overflow and enables an algorithmic stability (see Appendix A for details).

The series approximation \eqref{eq_12} alone covers the domain $0 < x < 40,000$ and ${10^{ - 4}} < y < {10^2}$, required in applications using the HITRAN molecular spectroscopic database \cite{Rothman2013}. In general, it provides accurate results while $y \geqslant {10^{ - 6}}$. However, this approximation may be used only to cover a smaller domain $0 \leqslant x \leqslant 15$ and ${10^{ - 6}} \leqslant y \leqslant 15$ that is considered most difficult for rapid and accurate computation.

In our recent publication we have shown that the following approximation (see equation \eqref{eq_6} in \cite{Abrarov2015a}) can be effective for computation in the narrow domain $0 \leqslant x \leqslant 15$ and $0 \leqslant y < {10^{ - 6}}$ along $x$-axis:
$$
\begin{aligned}
K\left( {x,y <  < 1} \right) &= \operatorname{Re} \left[ {w\left( {x,y <  < 1} \right)} \right] \\
   &\approx \operatorname{Re} \left\{ {{e^{{{\left( {ix - y} \right)}^2}}}\left[ {1 + \frac{{i{e^{{x^2}}}}}{{\sqrt \pi  }}\left( {2F\left( x \right) - \frac{{1 - {e^{2ixy}}}}{x}} \right)} \right]} \right\}
\end{aligned}
$$
or
\small
$$
K\left( {x,y <  < 1} \right) \approx {e^{{y^2} - {x^2}}}\cos \left( {2xy} \right) - \frac{{2{e^{{y^2}}}}}{{\sqrt \pi  }}\left[ {y\,{\rm{sinc}}\left( {2xy} \right) - F\left( x \right)\sin \left( {2xy} \right)} \right],
$$
\normalsize
where
$$
F\left( x \right) = {e^{ - {x^2}}}\int\limits_0^x {{e^{{t^2}}}} dt
$$
is the Dawson\text{'}s integral. As argument $x$ in the Dawson\text{'}s integral is real, its implementation is not problematic and several efficient approximations can be found in literature \cite{Cody1970, McCabe1974, Rybicki1989}.

When the input parameters $x$ and $y$ are large enough (say when the condition $\left| {x + iy} \right| > 15$ is satisfied), many rational approximations become effective for accurate and rapid computation. For example, the Gauss--Hermit quadrature or the Taylor expansion can be effectively implemented (see for example \cite{Letchworth2007} for details).

A Matlab source code for computation of the Voigt/complex error function that covers the entire complex plane can be accessed through Matlab Central, file ID: {\#}47801 \cite{Matlab2014}. This code has been developed by our research group and can be used for verification of the computed results. The domain divisions for computation of the Voigt function with complete coverage of the complex plane can be developed similarly.

In order to demonstrate the computational efficiency of the series approximation (12), the comparison with the Weideman\text{'}s rational approximation has been made (see equation (38-I) and corresponding Matlab code in \cite{Weideman1994}). Such a choice is justified since the Weideman\text{'}s approximation is one of the most rapid for computation of the Voigt/complex error function. The computational testing we performed by using a typical desktop computer shows that with same number of the summation terms ${m_{\max }} = 16$ (default integer in Matlab code in \cite{Weideman1994} is also $16$), the algorithm based on series approximation \eqref{eq_12} is faster in computation than that of based in the Weideman\text{'}s rational approximation by factors about $2.2$ and $2.7$ for input arrays $x$ and $y$ consisting of $5$ and $50$ million elements, respectively (see the Matlab source code with implementation of the series approximation \eqref{eq_12} in Appendix B). This is mainly because the Weideman\text{'}s rational approximation computes simultaneously both the real $K\left( {x,y} \right)$ and imaginary $L\left( {x,y} \right)$ parts, while the rational approximation \eqref{eq_12} computes only the real part $K\left( {x,y} \right)$ of the complex error function $w\left( {x,y} \right)$. It should be noted that in most practical applications the imaginary part $L\left( {x,y} \right)$ \eqref{eq_3} of the complex error function is not needed and simply ignored. Moreover, due to rapid convergence of the series approximation \eqref{eq_12} we may decrease the number of the summation terms. In particular, at ${m_{\max }} = 12$ the computational acceleration of the Voigt function can be further gained by about $30\% $. Therefore, the application of the series approximation \eqref{eq_12} may be advantageous especially for intense computations with extended input arrays.

\section{Error analysis}

In order to quantify accuracy of the series approximation (12), it is convenient to define the relative error as
$$
\Delta  = \left| {\frac{{K\left( {x,y} \right) - {K_{ref.}}\left( {x,y} \right)}}{{{K_{ref.}}\left( {x,y} \right)}}} \right|,
$$
where ${K_{ref.}}\left( {x,y} \right)$ is the reference. The highly accurate reference values can be generated, for example, by using the Algorithm 680 \cite{Poppe1990a, Poppe1990b} or recently published Algorithm 916 \cite{Zaghloul2011}.

Figures 2a and 2b show the logarithm ${\log _{10}}\Delta $ of the relative error of the series approximation \eqref{eq_12} at $m_{max} = 16$. The domain required for coverage of the HITRAN molecular spectroscopic database is  $0 < x < 40,000$ and ${10^{ - 4}} < y < {10^2}$ \cite{Quine2002, Wells1999} while the domain $0 \leqslant x \leqslant 15$ and ${10^{ - 6}} \leqslant y \leqslant 15$ is the most difficult for accurate and rapid computation of the Voigt function. Therefore, we will consider the accuracy behavior within the HITRAN subdomain and narrow band domain $0 \leqslant x \leqslant 15 \cap {10^{ - 4}} \leqslant y \leqslant 15$ and $0 \leqslant x \leqslant 15 \cap {10^{ - 6}} \leqslant y \leqslant {10^{ - 4}}$ separately as shown in Figs. 2a and 2b, respectively.

As we can see from Fig. 2a, within the HITRAN subdomain the accuracy of the series approximation is quite uniform and better than ${10^{ - 14}}$ over most of this area. Although the accuracy deteriorates with decreasing $y$, it, nevertheless, remains high and better than ${10^{ - 9}}$. Another advantage is that the area where the accuracy deteriorates is relatively small. Particularly, the area where accuracy is worse than ${10^{ - 13}}$ (yellow and red colors) is smaller than 2\% of the domain\text{'}s total area. 

\begin{figure}[ht]
\begin{center}
\includegraphics[width=32pc]{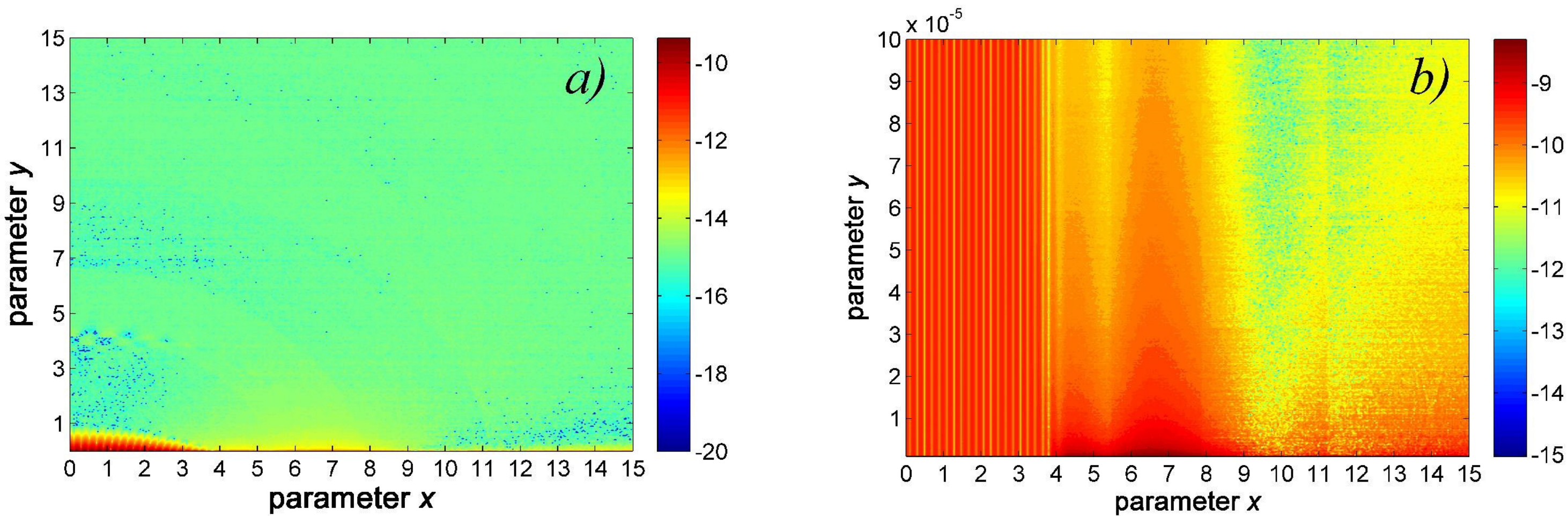}\hspace{2pc}%
\begin{minipage}[b]{28pc}
\vspace{0.3cm}
{\sffamily {\bf{Fig. 2.}} Logarithms of the relative error ${\log _{10}}\Delta $ for: a) for the HITRAN subdomain $0 \leqslant x \leqslant 15 \cap {10^{ - 4}} \leqslant y \leqslant 15$ and b) for the narrow band domain $0 \leqslant x \leqslant 15 \cap {10^{ - 6}} \leqslant y \leqslant {10^{ - 4}}$. The constants applied in computation are $\varsigma  = 2.75$, $N = 23$, ${m_{\max }} = 16$ and $h = 0.25$.}
\end{minipage}
\end{center}
\end{figure}

With randomly taken input parameters $x$ and $y$, it is determined that the average accuracy over the domain of practical interest $0 < x < 40,000$ and ${10^{ - 4}} < y < {10^2}$  is  ${10^{ - 14}}$. Although the series approximation \eqref{eq_12} can cover this domain accurately, it may be implemented only within domain $0 \leqslant x \leqslant 15$ and ${10^{ - 6}} \leqslant y \leqslant 15$ that is the most difficult for accurate and rapid computation.

In the narrow band shown in the Fig. 2b, the accuracy deteriorates further with decreasing $y$. However, it still remains high and better than ${10^{ - 8}}$. In particular, the best and worst accuracies in the narrow band domain $0 \leqslant x \leqslant 15 \cap {10^{ - 6}} \leqslant y \leqslant {10^{ - 4}}$ exceed ${10^{ - 10}}$ (yellow color) and ${10^{ - 8}}$ (dark red color), respectively.	

In modern applications requiring the HITRAN molecular spectroscopic data-base, the accuracy of the Voigt function should be ${10^{ - 6}}$. Therefore, we may reduce the integer ${m_{\max }}$ in the series approximation \eqref{eq_12} from $16$ to $12$ in order to gain computational acceleration. The number of the summation terms, determined by the integer ${m_{\max }}$, is quite sensitive to the small parameter value $h$. We have found empirically that at ${m_{\max }} = 12$ the best accuracy can be achieved by taking $h = 0.293$.

Figure 3a depicts the logarithm ${\log _{10}}\Delta $ of the relative error of the series approximation \eqref{eq_12} at $m_{max} = 12$ in the HITRAN subdomain $0 \leqslant x \leqslant 15$ and ${10^{ - 4}} \leqslant y \leqslant 15$. One can see that in the HITRAN subdomain the accuracy is better than ${10^{ - 8}}$.

Figure 3b illustrates the logarithm ${\log _{10}}\Delta $ of the relative error of the series approximation \eqref{eq_12} at $m_{max} = 12$ in the narrow band domain $0 \leqslant x \leqslant 15$ and ${10^{ - 6}} \leqslant y \leqslant {10^{ - 4}}$. Despite only $12$ summation terms involved in the series approximation (12), the accuracy within this domain is better than ${10^{ - 6}}$. For comparison, to achieve the same accuracy ${10^{ - 6}}$ at $y \geqslant {10^{ - 5}}$, the Weideman\text{'}s approximation requires $32$ summation terms (see Fig. 4 in \cite{Abrarov2011} for details). Thus, we can see that the series approximation \eqref{eq_12} may be useful and convenient in spectroscopic applications.

\begin{figure}[ht]
\begin{center}
\includegraphics[width=32pc]{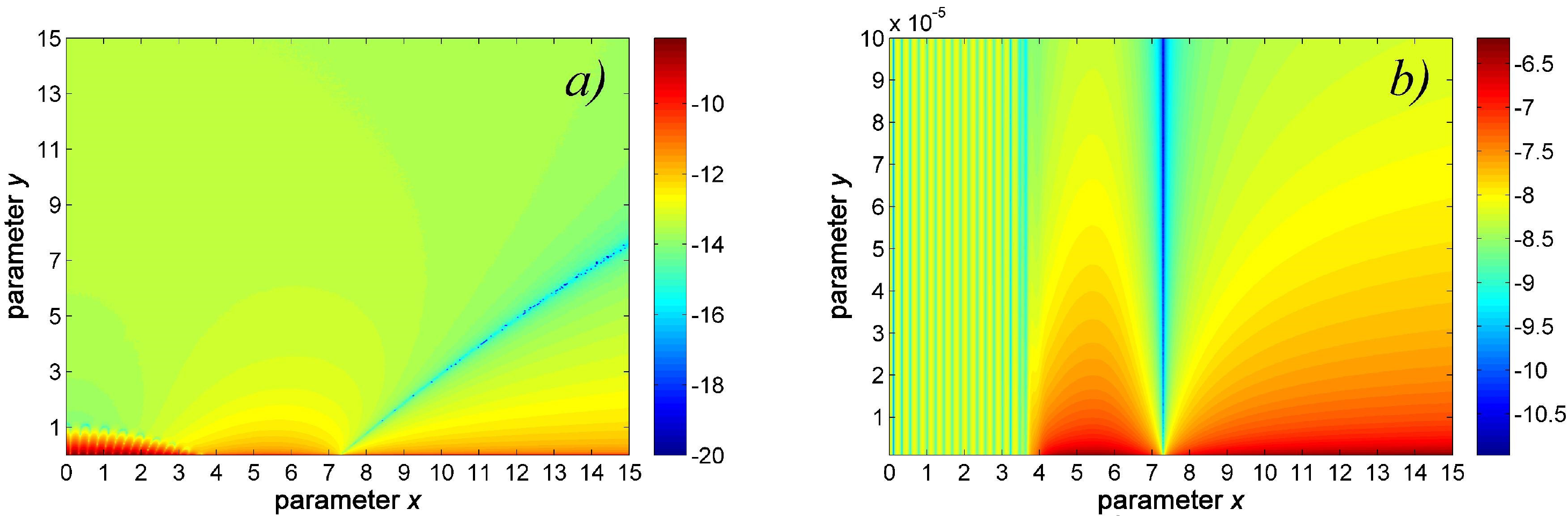}\hspace{2pc}%
\begin{minipage}[b]{28pc}
\vspace{0.3cm}
{\sffamily {\bf{Fig. 3.}} Logarithms of the relative error ${\log _{10}}\Delta $: a) for the HITRAN subdomain $0 \leqslant x \leqslant 15 \cap {10^{ - 4}} \leqslant y \leqslant 15$ and b) for the narrow band domain $0 \leqslant x \leqslant 15 \cap {10^{ - 6}} \leqslant y \leqslant {10^{ - 4}}$. The constants applied in computation are $\varsigma  = 2.75$, $N = 23$, ${m_{\max }} = 12$ and $h = 0.293$.}
\end{minipage}
\end{center}
\end{figure}

\section{Conclusion}

A rational approximation for rapid and accurate computation of the Voigt function is presented. With only $16$ summation terms, the proposed rational approximation provides average accuracy ${10^{ - 14}}$ in the domain of practical interest  $0 < x < 40,000$ and ${10^{ - 4}} < y < {10^2}$ that is needed for applications using the HITRAN molecular spectroscopic database. The computational test shows that the algorithm based on series approximation \eqref{eq_12} is more rapid in computation than that of based on the Weideman\text{'}s rational approximation by factor greater than $2$. Algorithmic stability is achieved since the proposed series approximation \eqref{eq_12} contains no poles at $y \geqslant 0$ and $ - \infty  < x < \infty $.

\section*{Acknowledgments}

This work is supported by National Research Council Canada, Thoth Technology Inc. and York University. The authors wish to thank to Prof. Ian McDade and Dr. Brian Solheim for discussions and constructive suggestions.

\section*{Appendix A}

According to definition of the $\kappa$-function \eqref{eq_12} we can write the following identity
\footnotesize
\[\label{eq_A.1}
\tag{A.1}
\begin{aligned}
&\kappa \left( {x,y + \varsigma /2} \right) \equiv \\
&\sum\limits_{m = 1}^{{m_{\max }}} {\frac{{{A_m}\left[ {C_m^2 - {x^2} + {{\left( {y + \varsigma /2} \right)}^2}} \right] + i{B_m}\left( {y + \varsigma /2} \right)\left[ {C_m^2 + {x^2} + {{\left( {y + \varsigma /2} \right)}^2}} \right]}}{{\left[ {{C_m} + x - i\left( {y + \varsigma /2} \right)} \right]\left[ {{C_m} - x + i\left( {y + \varsigma /2} \right)} \right]\left[ {C_m^2 - {{\left( {x + i\left( {y + \varsigma /2} \right)} \right)}^2}} \right]}}}
\end{aligned}
\]
\normalsize
and since
\scriptsize
\[\label{eq_A.2}
\tag{A.2}
\begin{aligned}
\left[ {{C_m} + x - i\left( {y + \varsigma /2} \right)} \right]\left[ {{C_m} - x + i\left( {y + \varsigma /2} \right)} \right]&\left[ {C_m^2 - {{\left( {x + i\left( {y + \varsigma /2} \right)} \right)}^2}} \right] \equiv \\
&\beta _m^2 + 2{\beta _m}\left( {{{\left( {y + \varsigma /2} \right)}^2} - {x^2}} \right) + {\left( {{x^2} + {{\left( {y + \varsigma /2} \right)}^2}} \right)^2},
\end{aligned}
\]
\normalsize
where the right side of the identity \eqref{eq_A.2} is the ${m^{{\text{th}}}}$ denominator of $\kappa \left( {x,}\right.$ $\left. {y + \varsigma /2} \right)$, it is sufficient to show that the poles do not exist on the right side of the identity (A.1) when both variables $x$ and $y$ are real such that $y \geqslant 0$ in order to prove that $\kappa \left( {x,y + \varsigma /2} \right)$ has no poles under same conditions.

The proof is not difficult. Let us equate the left side of identity \eqref{eq_A.2} to zero
\footnotesize
\[\label{eq_A.3}
\tag{A.3}
\left[ {{C_m} + x - i\left( {y + \varsigma /2} \right)} \right]\left[ {{C_m} - x + i\left( {y + \varsigma /2} \right)} \right]\left[ {C_m^2 - {{\left( {x + i\left( {y + \varsigma /2} \right)} \right)}^2}} \right] = 0.
\]
\normalsize
and then solve this equation with respect to the variables $x$ and $y$. Suppose now that the solutions in the equation \eqref{eq_A.3} for real valued arguments $x \in \left( { - \infty ,\infty } \right)$ and $y \in \left[ {0,\infty } \right)$ exist.  Solving the equation \eqref{eq_A.3} with respect to $x$ results in four possible solutions ${x_1} =  - i\left( {y + \varsigma /2} \right) - {C_m}$, ${x_2} = i\left( {y + \varsigma /2} \right) + {C_m}$, ${x_3} =  - i\left( {y + \varsigma /2} \right) + {C_m}$ and ${x_4} = i\left( {y + \varsigma /2} \right) - {C_m}$. Since the constants ${C_m}$, $\varsigma $ are real valued and since $\varsigma  > 0$, $y \geqslant 0$, these solutions $\left\{ {{x_1},{x_2},{x_3},{x_4}} \right\}$ must be always complex. However, the complex solutions $\left\{ {{x_1},{x_2},{x_3},{x_4}} \right\}$ contradict our initial assumption that $x$ is real. Similarly, four possible solutions of equation \eqref{eq_A.3} with respect to the variable $y$ are ${y_1} =  - i\left( {x - {C_m}} \right) - \varsigma /2$, ${y_2} = i\left( {x - {C_m}} \right) - \varsigma /2$, ${y_3} =  - i\left( {x + {C_m}} \right) - \varsigma /2$ and ${y_4} = i\left( {x + {C_m}} \right) - \varsigma /2$. Since the constants ${C_m}$, $\varsigma $ are real and positive, the solutions $\left\{ {{y_1},{y_2},{y_3},{y_4}} \right\}$ must be either complex at $x \ne {C_m}$ or negative and equal to $ - \varsigma /2$ at $x = {C_m}$. However, the complex or negative solutions $\left\{ {{y_1},{y_2},{y_3},{y_4}} \right\}$ contradict our initial assumption that $y \geqslant 0$. Due to these contradictions we must conclude that there are no poles in identity \eqref{eq_A.3} under the conditions $\left\{ {x,y} \right\} \in  \mathbb{R}$ such that $y \geqslant 0$.

The absence of the poles signifies that while the arguments is taken by absolute value as $\left| y \right|$, the function $\kappa \left( {x,\left| y \right| + \varsigma /2} \right)$ will never encounter division to zero that leads to computational overflow. That is why taking the input parameter by absolute value as $\left| y \right|$ is advantageous since this approach provides stable performance of the algorithm.

\section*{Appendix B}
\footnotesize
\begin{verbatim}
function VF = voigtf(x,y,opt)
 
% This function file is a subroutine for computation of the Voigt function.
% The input parameter y is used by absolute value according to the
% procedure described in the article. The parameter opt is either 1 for 
% more accurate or 2 for more rapid computation. At y < 0 change the sign  
% to negative externally, out of the body of this function file.
%
% NOTE: This program completely covers the domain 0 < x < 40,000 and 
% 10^-4 < y < 10^2 required for applications using the HITRAN molecular 
% spectroscopic database. However, it may be implemented only to cover the 
% smaller domain 0 <= x <= 15 and 10^-6 <= y <= 15 that is the most 
% difficult for rapid and accurate computation. See the article that 
% briefly describes how other domains can be covered.
%
% The code is written by Sanjar M. Abrarov and Brendan M. Quine, York
% University, Canada, March 2015.
 
if nargin == 2
    opt = 1;
end

if opt ~= 1 && opt ~=2
    disp(['opt = ',num2str(opt),' cannot be assigned. Use either 1 or 2.'])
    return
end

% *************************************************************************
% Define array of coefficients as coeff = [alpha;beta;gamma]'
% *************************************************************************
if opt == 1
    
    coeff = [
    1.608290174437121e-001 3.855314219175531e-002  1.366578214428949e+000
    6.885967427017463e-001 3.469782797257978e-001 -5.742919588559361e-002
    2.651151642675390e-001 9.638285547938826e-001 -5.709602545656873e-001
   -2.050008245317253e-001 1.889103967396010e+000 -2.011075414803758e-001
   -1.274551644219086e-001 3.122804517532180e+000  1.069871368716704e-002
   -1.134971805306579e-002 4.664930205202391e+000  1.468639542320982e-002
    4.201921570328543e-003 6.515481030406647e+000  1.816268776500938e-003
    8.084740485193432e-004 8.674456993144942e+000 -6.875907999947567e-005
    1.946391440605860e-005 1.114185809341728e+001 -2.327910355924500e-005
   -4.132639863292073e-006 1.391768433122366e+001 -1.004011418729134e-006
   -2.656262492217795e-007 1.700193570656409e+001  2.304990232059197e-008
   -1.524188131553777e-009 2.039461221943855e+001  2.275276345355270e-009
    2.239681784892829e-010 2.409571386984707e+001  3.383885053101652e-011
    4.939143128687883e-012 2.810524065778962e+001 -4.398940326332977e-013
    4.692078138494072e-015 3.242319258326621e+001 -1.405511706545786e-014
   -2.512454984032184e-016 3.704956964627684e+001 -3.954682293307548e-016
    ]; 
    mMax = 16; % 16 summation terms
 
elseif opt == 2
        
    coeff = [
    2.307372754308023e-001 4.989787261063716e-002  1.464495070025765e+000
    7.760531995854886e-001 4.490808534957343e-001 -3.230894193031240e-001
    4.235506885098250e-002 1.247446815265929e+000 -5.397724160374686e-001
   -2.340509255269456e-001 2.444995757921221e+000 -6.547649406082363e-002
   -4.557204758971222e-002 4.041727681461610e+000  2.411056013969393e-002
    5.043797125559205e-003 6.037642585887094e+000  4.001198804719684e-003
    1.180179737805654e-003 8.432740471197681e+000 -5.387428751666454e-005
    1.754770213650354e-005 1.122702133739336e+001 -2.451992671326258e-005
   -3.325020499631893e-006 1.442048518447414e+001 -5.400164289522879e-007
   -9.375402319079375e-008 1.801313201244001e+001  1.771556420016014e-008
    8.034651067438904e-010 2.200496182129099e+001  4.940360170163906e-010
    3.355455275373310e-011 2.639597461102705e+001  5.674096644030151e-014
    ];
    mMax = 12; % 12 summation terms
end
% *************************************************************************

varsigma = 2.75; % define the shift constant
y = abs(y) + varsigma/2;

arr1 = y.^2 - x.^2; % define 1st repeating array
arr2 = x.^2 + y.^2; % define 2nd repeating array
arr3 = arr2.^2;  % define 3rd repeating array
 
    VF = 0; % initiate VF
    for m = 1:mMax
        VF = VF + (coeff(m,1)*(coeff(m,2) + arr1) + ...
            coeff(m,3)*y.*(coeff(m,2) + arr2))./(coeff(m,2)^2 + ...
            2*coeff(m,2)*arr1 + arr3);
    end
end

\end{verbatim}
\normalsize


\end{document}